\documentclass[11pt,a4paper]{article}
\usepackage[utf8]{inputenc}
\usepackage[T1]{fontenc}
\usepackage[normalem]{ulem} 
\usepackage{amsmath}
\usepackage{amsfonts}
\usepackage{amssymb}
\usepackage{amsthm}
\usepackage{mathtools}
\usepackage{bbm}
\usepackage{setspace}
\usepackage{graphicx}
\usepackage{hyperref}
\usepackage{tikz}
\hypersetup{
        citecolor=black,
	colorlinks=true,
	linkcolor=black,
	filecolor=black,
	urlcolor=black,
}
\usepackage{cleveref}
\usepackage{natbib}
\usepackage{soul}
\usepackage{float}
\usepackage{verbatim}
\usepackage{longtable,booktabs}
\usepackage[font=footnotesize,labelfont=bf]{caption}
\usepackage{subcaption} 
\usepackage{makecell}
\usepackage{xcolor}
\usepackage[margin=0.75in]{geometry}
\usepackage{xr}
\usepackage{caption}
\usepackage{subcaption}
\usepackage{color}

\renewcommand{\baselinestretch}{1}

\author{Rui Feng\footnote{Feng is PhD student, Department of Statistics, University of Warwick, Email: rui.feng.1@warwick.ac.uk} 
 \and  Chenlei Leng\footnote{Leng is Professor, Department of Statistics, University of Warwick, Email: c.leng@warwick.ac.uk. Corresponding author.}}
\title{Modelling Directed Networks with Reciprocity}
\date{}
\providecommand{\keywords}[1]
{
	\small	
	\textbf{\textit{Key words:}} #1
}

\newtheorem{theorem}{Theorem}
\newtheorem{Prop}{Proposition}
\newtheorem{Lem}{Lemma}
\newtheorem{corollary}{Corollary}
	
\theoremstyle{definition}

\newtheorem{Assum}{Assumption}

\newtheorem*{Prop*}{Proposition}
\newtheorem*{theorem*}{Theorem}

\begin{document}
\maketitle
	
\begin{abstract}
\begin{singlespace}

Asymmetric relational data is increasingly prevalent across diverse fields, underscoring the need for directed network models to address the complex challenges posed by their unique structures. Unlike undirected models, directed models can capture reciprocity, the tendency of nodes to form mutual links. In this work, we address a fundamental question: what is the effective sample size for modeling reciprocity? We examine this by analyzing the Bernoulli model with reciprocity, allowing for varying sparsity levels between non-reciprocal and reciprocal effects. We then extend this framework to a model that incorporates node-specific heterogeneity and link-specific reciprocity using covariates. Our findings reveal intriguing interplays between non-reciprocal and reciprocal effects in sparse networks. We propose a straightforward inference procedure based on maximum likelihood estimation that operates without prior knowledge of sparsity levels, whether covariates are included or not.
	
\end{singlespace}
\end{abstract}


\section{Introduction}
Consider a directed network with \( n \) nodes, denoted by \( G_n = (V, E) \), where \( V = \{1, \dots, n\} \) is the set of nodes and \( E \subseteq V \times V \) represents the edge set. We focus on simple graphs, so no self-loops are allowed, i.e., \( (j,j) \notin E \) for any \( j \in V \). Let \( A_{ij} \in \{0, 1\} \) denote the random variable indicating the presence of a directed link from node \( i \) to node \( j \). Assuming that dyads \( (A_{ij}, A_{ji}) \) and \( (A_{kl}, A_{lk}) \) are independent whenever \( \{i,j\} \cap \{k,l\}=\emptyset \), the Bernoulli model with reciprocity (BR) specifies multinomial probabilities for each dyad as follows \citep{krivitsky2015question}:
\begin{equation}\label{eq:BR}
\textbf{BR model:} \quad
p_{ij}(0,0)  \propto 1, \quad 
p_{ij}(1,0) = p_{ij}(0,1) \propto \exp(\mu_n), \quad 
p_{ij}(1,1) \propto \exp(2 \mu_n + \rho_n),
\end{equation}
where \( p_{ij}(a, b) = p(A_{ij} = a, A_{ji} = b) \). 

In this model, \( \mu_n \) represents the baseline tendency of nodes \( i \) and \( j \) to connect, while \( \rho_n \) captures \textit{reciprocity}, the propensity for pairs of nodes to form mutual links. A positive \( \rho_n \) suggests that reciprocal ties occur more frequently than would be expected if all links were independent, whereas a negative \( \rho_n \) indicates a tendency to avoid forming mutual links. This model raises a fundamental question:
\begin{center}
    \textit{Question 1: What is the effective sample size for the statistical inference of \( \mu_n \) and \( \rho_n \)?}
\end{center}

This question would be straightforward if \( \mu_n \) and \( \rho_n \) were both fixed, as it would fall under standard maximum likelihood estimation. However, this paper addresses the question when \( \mu_n \) and \( \rho_n \) vary with \( n \) in a single-network asymptotic framework, allowing \( n \), the number of nodes, to grow indefinitely. Particularly relevant is the regime where the network is sparse, meaning that \( \sum_{i,j} p_{ij} = o(n^2) \), so the total number of links grows at a slower rate than the maximum possible number of connections.

\subsection{A framework for sparse networks}
A discerning reader may wonder why the BR model is of particular interest. There are several reasons. First, this model serves as a natural extension of the Erd\H{o}s--R\'enyi model \citep{renyi1959random,erdHos1960evolution} for undirected graphs, adapted to incorporate reciprocity for the analysis of directed networks. The Erd\H{o}s--R\'enyi model is foundational in network science, underpinning nearly all modern stochastic models for graphs, as discussed in the literature review below. In the Erd\H{o}s--R\'enyi model, links are symmetric (\(A_{ij} = A_{ji}\)) and form independently across dyads with probability given by \(\text{logit}\left(p(A_{ij} = 1)\right) = \mu_n\), where \(\mu_n\) serves as a density parameter. The scaling \(\mu_n \asymp -\log(n)\) is of particular interest, as it determines various structural properties of a realized graph, such as the emergence of a giant component.

The effective sample size for inferring \(\mu_n\) in the Erd\H{o}s--R\'enyi  model has been fully explored under the regime \(\mu_n = -a\log(n) + \mu\), where \(\mu\) and \(a\) are fixed constants \citep{chen2021analysis}. However, for the BR model, the inference of both \(\mu_n\) and \(\rho_n\) has only been partially addressed in \cite{krivitsky2015question}, which requires the effective sample sizes for these parameters to be of the same order. Extending the analysis to allow different sparsity levels for \(\mu_n\) and \(\rho_n\) addresses Question $1$ in a more comprehensive way, offering insights into the effective sample sizes required for a wider range of network structures.

More importantly, a complete answer to this question will pave the way for developing new models. As an example, we extend the BR model to the following:
\begin{align}\label{eq:p1.5}
\mathbf{p_{1.5}}~\textbf{model}: \quad &p_{i j}(0,0) \propto 1, \quad p_{ij}(1,0) \propto \exp \left(\mu_n+X_{i}^{T} \gamma_1+Y_{j}^{T} \gamma_2\right), \nonumber\\  
&p_{ij}(0,1) \propto \exp \left(\mu_n+X_{j}^{T} \gamma_1+Y_{i}^{T}  \gamma_2\right), \nonumber\\
&p_{ij}(1,1) \propto \exp \left(2\mu_n+\left(X_{i}^{T}+X_{j}^{T}\right) \gamma_1+\left(Y_{i}^{T}+Y_{j}^{T}\right) \gamma_2+\rho_n+V_{i j}^{T} \delta\right),
\end{align}
with additional parameters \(\gamma_1\), \(\gamma_2\), and \(\delta\), where \(X_i \in \mathbb{R}^{d_1}\) represents covariates related to node \(i\)'s outgoingness, \(Y_i\in\mathbb{R}^{d_2}\) relates to its incomingness, and \(V_{ij}\in\mathbb{R}^{d_3}\) governs the reciprocity between nodes \(i\) and \(j\). The model in \eqref{eq:p1.5} allows for node-specific heterogeneity via \(X_i^T\gamma_1\) for outgoingness and \(Y_j^T\gamma_2\) for incomingness, and \(V_{ij}^{T} \delta\) to model heterogeneity in reciprocal relationships. Assuming that the parameters associated with the covariates are fixed, we further pose the following question:

\begin{center}
\textit{Question 2: What are the effective sample sizes for the statistical inference of \(\gamma_1\), \(\gamma_2\), and \(\delta\)?}
\end{center}

The model in \eqref{eq:p1.5} has a close relationship with the \(p_1\)  model introduced by \cite{holland1981exponential}, which is specified as follows:
\begin{align*}
\mathbf{p_1}~\textbf{model}: \quad &p_{i j}(0,0) \propto 1, \quad p_{i j}(1,0) \propto \exp \left(\mu+\alpha_i+\beta_j\right), \\
&p_{i j}(0,1) \propto \exp \left(\mu+\alpha_j+\beta_i\right), \\
&p_{i j}(1,1) \propto \exp \left(2 \mu+\rho+\alpha_i+\alpha_j+\beta_i+\beta_j\right),
\end{align*}
where the \(p_1\) model employs node-specific fixed effects without explicitly accounting for link-specific reciprocity. Our model in \eqref{eq:p1.5} parametrizes these fixed effects through covariates, achieving a more parsimonious structure. Although it may lack some of the flexibility of the \(p_1\) model, this approach offers certain advantages, such as enabling link prediction for new nodes not used in model fitting. Additionally, a key advantage of the model in \eqref{eq:p1.5} lies in its suitability for sparser networks. We show that inference is feasible as long as the number of links diverges. In contrast, the \(p_1\) model, with its large number of parameters, typically requires much denser networks to ensure the existence and asymptotic normality of its estimators, though no formal inference procedures are currently available for these estimators (see literature review below). Additionally, the model in \eqref{eq:p1.5} shares features with the \(p_2\) model \citep{van2004p2}, which also includes random effects for outgoingness and incomingness. As our model conceptually bridges the \(p_1\) and \(p_2\) models, we refer to it as the \(p_{1.5}\) model.

\subsection{Literature review}
A substantial portion of modern datasets reflects relational structures, where data capture relationships between entities. This paper focuses on asymmetric relationship--those with a distinct directionality. Examples include followee-follower relations on social media, citee-citer relationships in academic publications, import-export dependencies in economics, and predator-prey interactions in ecology. These relationships are naturally represented as directed networks, with nodes depicting entities and directed links denoting the relationships. Developing models to capture the generative mechanisms of these structures is essential for addressing key questions, such as why and how directed ties form, what factors drive their formation, and how network structures can inform decision-making processes.

Due to the extensive prevalence of relational data across various disciplines, network analysis has emerged as a dynamic, multidisciplinary field encompassing statistics, applied mathematics, economics, social science, biology, medicine, neuroscience, engineering, and more. As relational data has expanded, so too has the development of statistical models for analyzing network structures. A substantial body of models, primarily focused on undirected networks and often derived from or extending the foundational Erd\H{o}s--R\'enyi model, has been established in this domain. Notable examples include the stochastic block model \citep{holland1983stochastic}, graphon models \citep{bickel2009nonparametric,wolfe2013nonparametric}, latent space models \citep{hoff2002latent, ma2020universal}, the $\beta$-model \citep{chatterjee2011random,chen2021analysis}, and exponential random graph models \citep{robins2007introduction}.

In contrast, systematic studies of directed networks are less common. Important early contributions focused on algebraic properties and maximum likelihood estimators within the \( p_1 \) model framework, as explored by \cite{petrovic2010algebraic} and \cite{rinaldo2010existence}. Further developments by \cite{yan2016asymptotics,yan2019statistical} investigated the asymptotic properties of the maximum likelihood estimator for the \( p_1 \) model, while \cite{qu2024inference} considered a semiparametric model with an unknown link function. None of these papers considered incorporating reciprocity effects. On the other hand, when reciprocity is present, existing studies have relied on numerical justifications rather than theoretical results \citep{yan2015simulation}.

The remainder of this paper is organized as follows. Section \ref{Sec: toy} provides a general analysis of the Bernoulli model. Section \ref{Sec: Main results} introduces the \( p_{1.5} \) model and presents asymptotic results. Section \ref{Sec: Simulation} details our simulation results, and Section \ref{Sec: Data Analysis} demonstrates our model's application on two real datasets. 
All proofs are provided in the Supplementary Materials.

In this paper, a subscript \( n \) on a parameter (e.g., \( \mu_n \), \( \rho_n \)) indicates its dependence on \( n \). Parameters without this subscript are independent of \( n \). True parameter values are denoted by a subscript \( 0 \), such as \( \mu_{n0} \) for the true value of \( \mu_n \) and \( \gamma_{10} \) for the true value of \( \gamma_{1} \).

\section{The Bernoulli Model with Reciprocity}\label{Sec: toy}
We begin by examining the effective sample sizes for the Bernoulli model (BR model) as specified in \eqref{eq:BR}. For the sake of theoretical analysis and notational convenience, it is beneficial to work with the parameters \((\mu_n, \tau_n)\), where \(\tau_n = 2\mu_n + \rho_n\). The negative log-likelihood function with respect to \((\mu_n, \tau_n)\) can be expressed as:
\[
\ell_{n}^{(1)}(\mu_n, \tau_n) = \sum_{i<j} \log(k_{n,ij}) - \mu_n \sum_{i<j} \left( A_{ij}(1 - A_{ji}) + A_{ji}(1 - A_{ij}) \right) - \tau_n \sum_{i<j} A_{ij} A_{ji},
\]
where \(k_{n,ij} = 1 + 2 \exp(\mu_n) + \exp(\tau_n)\) serves as the normalizing constant. 
It is important to note that the likelihood functions defined in terms of \((\mu_n, \rho_n)\) and \((\mu_n, \tau_n)\) are equivalent, as are their corresponding maximum likelihood estimators. This leads us to the following lemma:

\begin{Lem}\label{lemma1}
Suppose \((\hat{\mu}_n, \hat{\tau}_n) = \text{argmin}_{(\mu_n, \tau_n) \in \mathbb{R}^2} ~\ell_{n}^{(1)}(\mu_n, \tau_n)\). Then, it follows that \\\((\hat{\mu}_n, \hat{\tau}_n - 2\hat{\mu}_n) = \text{argmin}_{(\mu_n, \rho_n) \in \mathbb{R}^2} ~\ell_{n}^{(2)}(\mu_n, \rho_n)\), where \(\ell_{n}^{(2)}(\mu_n, \rho_n)\) denotes the negative log-likelihood function parametrized by \(\mu_n\) and \(\rho_n\). The reverse direction also holds.
\end{Lem}

Given the equivalence between the likelihood functions, we will focus on estimating \(\mu_n\) and \(\tau_n\) in the subsequent analysis. Drawing inspiration from the significant role that \(-\log n\) plays in the Erd\H{o}s--R\'enyi model for modeling sparse networks, we define 
\[
\mu_n = -a \log n + \mu, \quad \tau_n = -b \log n + \tau,
\]
where \(\mu \in \left[-M_\mu, M_\mu\right]\) and \(\tau \in \left[-M_\tau, M_\tau\right]\) are fixed constants with \(M_\mu\) and \(M_\tau\) specified, and without loss of generality, we assume \(a > 0\) and \(b > 0\). From \(\ell_{n}^{(1)}(\mu_n, \tau_n)\), we can interpret \(a\) as the sparsity index of non-reciprocal links, represented by \(\sum_{i<j} \left( A_{ij}(1 - A_{ji}) + A_{ji}(1 - A_{ij}) \right)\), while \(b\) serves as the sparsity index for reciprocal links, given by \(\sum_{i<j} A_{ij} A_{ji}\).

This transformation clarifies the dependence of sparsity on \(n\) while allowing for intuitive statistical inference on the fixed parameters \(\mu\) and \(\tau\). For further discussions on this topic, we refer to \cite{krivitsky2015question} and \cite{chen2021analysis}. It is important to note that the constants \(a\), \(\mu\), \(b\), and \(\tau\) are not identifiable or estimable. To address these challenges, we will later develop a straightforward inference procedure for \(\mu_n\) and \(\tau_n\).

Under the given scaling, we find that the expected number of non-reciprocal links is
\[
\mathbb{E}\left(\sum_{{i,j}=1}^n A_{ij} - \sum_{i<j} A_{ij} A_{ji}\right) \asymp n^{2-a},
\]
while the expected number of reciprocated links is
\[
\mathbb{E}\left(\sum_{i<j} A_{ij} A_{ji}\right) \asymp n^{2-b}.
\]
Consequently, the total expected number of links is of order \(n^{2-a}\) if \(a \le b\), or \(n^{2-b}\) if \(a > b\). This scaling choice highlights that the two quantities can indeed differ in magnitude. Notably, \cite{krivitsky2015question} examined a special case of our framework when \(a = b = 1\), leading to comparable expected numbers of non-reciprocal and reciprocated links.

We now derive the effective sample sizes for \(\mu\) and \(\tau\), assuming that \(a\) and \(b\) are known. We begin by expressing the negative log-likelihood function as follows:
\begin{align}\label{NLF}
\ell_n(\mu, \tau) = \sum_{i<j} \log(k_{ij}) - \mu \sum_{i<j} \left( A_{ij}(1 - A_{ji}) + A_{ji}(1 - A_{ij}) \right) - \tau \sum_{i<j} A_{ij} A_{ji},
\end{align}
where \(k_{ij} = 1 + 2 n^{-a} \exp(\mu) + n^{-b} \exp(\tau)\) serves as the normalizing constant. Our maximum likelihood estimator is defined as 
\[
(\hat{\mu}, \hat{\tau}) = \operatorname{argmin}_{(\mu,\tau) \in \Omega_1} \frac{1}{\binom{n}{2}} \ell_n(\mu, \tau),
\]
with \(\Omega_1 \triangleq \left[-M_\mu, M_\mu\right] \times \left[-M_\tau, M_\tau\right]\). To derive the asymptotic results, we make the following assumptions:

\begin{Assum}{\label{ass1}} 
Assume \(0 < a, b < 2\). The true values \((\mu_{0}, \tau_{0})\) lie within the interior of \(\Omega_1\).
\end{Assum}

The conditions \(a > 0\) and \(b > 0\) ensure that the resulting graph is sparse, while \(a < 2\) and \(b < 2\) are necessary to guarantee that the total numbers of reciprocal and non-reciprocal links approach infinity. Without these conditions, consistent estimation would not be achievable. We now present the following result regarding the maximum likelihood estimator (MLE). All our results hold under Assumption \ref{ass1}, meaning they apply to arbitrarily sparse networks.

\begin{Prop}{\label{prop1}}
Under Assumption \ref{ass1}, as \(n\) approaches infinity, the MLE \((\hat{\mu}, \hat{\tau})\) is consistent and asymptotically normal, specifically:
\[
\left( \sqrt{n^{2-a}}(\hat{\mu} - \mu_{0}), \quad \sqrt{n^{2-b}}(\hat{\tau} - \tau_{0}) \right)^T \stackrel{d}{\longrightarrow} N(0, \Sigma^{-1}),
\]
where 
\[
\Sigma = \begin{pmatrix} \exp(\mu_{0}) & 0 \\ 0 & \exp(\tau_{0})/2 \end{pmatrix}.
\]
\end{Prop}

Following the reasoning in \cite{krivitsky2015question} and \cite{chen2021analysis}, we can interpret \(n^{1-a/2}\) and \(n^{1-b/2}\) as the effective sample sizes for \(\mu\) and \(\tau\), respectively. This interpretation is intuitive, as from equation \eqref{NLF}, \(\mu\) can be seen as the density parameter for the configuration \((1,0)\) and \((0,1)\), while \(\tau\) represents the density parameter for the configuration \((1,1)\).

Next, we turn our attention to the estimation of \(\rho_n\). We define \(\hat{\rho} = \hat{\tau} - 2\hat{\mu}\) and \(\rho_0 = \tau_0 - 2\mu_0\). Based on Proposition \ref{prop1}, we derive the following corollary:

\begin{corollary} \label{rho}
Under Assumption \ref{ass1}, as \(n\) approaches infinity, we have $\sqrt{n^{2-\max\{a,b\}}} \left( \hat{\rho} - \rho_0 \right) \xrightarrow{d} N(0,v)$ where $v$ is the asymptotic variance and varies from different scenarios. Specifically, if $a = b$, $v=2\exp^{-1}(2 \mu_0 + \rho_0)+4\exp^{-1}(\mu_0)$. If $a < b$,  $v=2\exp^{-1}(2 \mu_0 + \rho_0)$ and if $a > b$, $v=4\exp^{-1}(\mu_0)$.
\end{corollary}

Corollary \ref{rho} demonstrates that the convergence rate and limiting distribution of \(\hat{\rho}\) depend on the relative values of \(a\) and \(b\). Given that \(\rho\) is the reciprocity parameter influencing clique formation, it provides an additional comparative effect between the densities of the configurations \(\{(1,0),(0,1)\}\) and \((1,1)\). Consequently, the effective sample size for \(\hat{\rho}\) is determined by the smaller effective sample size between \(\hat{\mu}\) and \(\hat{\tau}\), which is \(O(n^{2 - \max\{a,b\}})\).

Notably, when \(a \ge b\), two distinct regimes arise for the asymptotic distribution of \(\hat{\rho}\). Specifically, when \(a > b\), the asymptotic variance is given by \(4\exp^{-1}(\mu_0)\), while if \(a = b\), an additional factor of \(2\exp^{-1}(2 \mu_0 + \rho_0)\) appears, reflecting the contribution of reciprocal ties to the asymptotic variance. Thus, Proposition \ref{prop1} and Corollary \ref{rho} provide a comprehensive answer to Question 1 regarding the effective sample sizes of \(\mu_n\) and \(\rho_n\) in the context of their statistical inference.

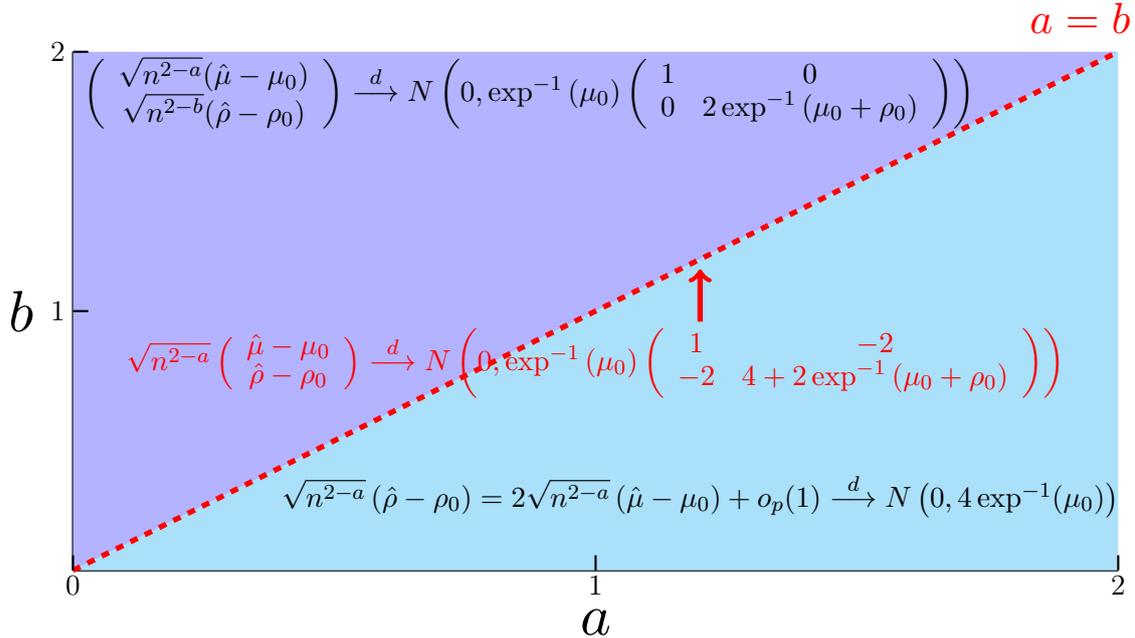
\begin{figure}[!htbp]
    \centering
\begin{tikzpicture}

  \message{sparsity level diagram}
  \def\tick{0.05*\xmax}
  \def\xmax{0.8*\textwidth}
  \def\ymax{0.4*\textwidth}
  \def\N{40}
  \def\isotherm#1#2{{ 1.6*#2/#1 }}
  
  \draw[-,thick] (0,0) -- (0,\ymax)
    node[pos=0.5,left,inner sep=7,scale=2] at (0, 0.5*\ymax) {$b$};
  \draw[-,thick] (0,0) -- (\xmax,0)
    node[pos=0.5,below,inner sep=7,scale=2,anchor=north] {$a$};

\fill[blue!30] (0,0) -- (\xmax,\ymax) -- (0,\ymax) -- cycle;

\fill[cyan!30] (0,0) -- (\xmax,0) -- (\xmax,\ymax) -- cycle;

 \draw[thick, red, dashed, line width=2pt] (0, 0) -- (\xmax, \ymax) node[above, xshift=-0.5cm,scale=1.5] {$a=b$};
 
 \node at  (6,0.92*\ymax) {$\left(\begin{array}{c}\sqrt{n^{2-a}}(\hat{\mu}-\mu_0)\\ \sqrt{n^{2-b}}(\hat{\rho}-\rho_0)\end{array}\right) \stackrel{d} {\longrightarrow} N\left(0,\exp ^{-1}\left(\mu_0\right) \bigg(\begin{array}{cc}
1& 0 \\
0& 2 \exp ^{-1}\left( \mu_0+\rho_0\right)
\end{array}\bigg)\right)$};
 \node[red] at  (0.5*\xmax,0.4*\ymax)  {$\sqrt{n^{2-a}}\left(\begin{array}{c}\hat{\mu}-\mu_0\\ \hat{\rho}-\rho_0\end{array}\right) \stackrel{d} {\longrightarrow} N\left(0,\exp ^{-1}\left(\mu_0\right) \bigg(\begin{array}{cc}
1& -2 \\
-2& 4+2 \exp ^{-1}\left( \mu_0+\rho_0\right)
\end{array}\bigg)\right)$};
\node at  (0.6*\xmax,1)  {$\sqrt{n^{2-a}}\left(\hat{\rho}-\rho_0\right)=2 \sqrt{n^{2-a}}\left(\hat{\mu}-\mu_0\right)+o_p(1) \stackrel{d} {\longrightarrow} N\left(0,4 \exp^{-1}(\mu_0)\right)$};
\node (A) at  (0.6*\xmax,0.46*\ymax) {};
\node (B) at  (0.6*\xmax,0.6*\ymax) {};

\draw[->,line width=2pt, red] (A) -- (B);

\foreach \x in {0,1,2} {
        \draw[thick] (\x*\xmax*0.5, 0) -- (\x*\xmax*0.5, 0.2); 
        \node at (\x*0.5*\xmax, -0.2) {\x};              
    }
\foreach \y in {1,2} {
        \draw[thick] (0,\ymax*0.5*\y) -- (0.2,\ymax*0.5*\y);  
         \node at (-0.2,\ymax*0.5*\y) {\y};             
    }
    
\end{tikzpicture}
    \caption{Asymptotic distributions illustrating the phase transition phenomenon driven by the sparsity indices $a$ and $b$. The dashed red line is the 45 degree line where $a=b$.}
    \label{dependence}
\end{figure}

Using Proposition \ref{prop1} and Corollary \ref{rho}, along with some straightforward calculations, we can determine the asymptotic distribution of $(\hat\mu - \mu_0, \hat\rho - \rho_0)$, which varies across three regimes based on the relative values of $a$ and $b$. Specifically, the joint asymptotic distribution exhibits a phase transition, influenced by the relative magnitudes of the sparsity indices $a$ and $b$, as illustrated in Figure \ref{dependence}. {Notably, in the asymptotic sense, the relationship between $\hat\mu$ and $\hat\rho$ transitions from complete independence when $a<b$ to perfect dependence with a correlation of one when $a>b$.
}

While Proposition \ref{prop1} and Corollary \ref{rho} provide valuable theoretical insights, it is important to note that, in practice, the sparsity level parameters \(a\) and \(b\) are typically unknown. This lack of information complicates the solution of equation (\ref{NLF}) and the estimation of the MLE \((\hat{\mu}, \hat{\tau})\). However, we can address this challenge using the following approach. Let \((\hat{\mu}_n, \hat{\tau}_n)\) represent the MLE of \(\ell^{(1)}_{n}(\mu_n, \tau_n)\) that does not depend on knowledge of \(a\) and \(b\). We define \(\hat{\rho}_n = \hat{\tau}_n - 2\hat{\mu}_n\). The following proposition summarizes our findings:

\begin{Prop}{\label{prop2}}  
Under Assumption \ref{ass1}, as \(n\) approaches infinity, the maximum likelihood estimates \(\hat{\mu}_n\) and \(\hat{\tau}_n\) exist with probability approaching one, and we have:
\begin{align*}
&n\sqrt{\exp(\hat{\mu}_n)}\left(\hat{\mu}_n - \mu_{n0}\right) \stackrel{d}{\longrightarrow} N(0, 1),\quad 
n\sqrt{\exp(\hat{\tau}_n)}\left(\hat{\tau}_n - \tau_{n0}\right) \stackrel{d}{\longrightarrow} N(0, 2),\\ 
&\frac{n\sqrt{\exp(\hat{\mu}_n)\exp(2\hat{\mu}_n + \hat{\rho}_n)}}{\sqrt{2\exp(\hat{\mu}_n) + 4\exp(2\hat{\mu}_n + \hat{\rho}_n)}}\left(\hat{\rho}_n - \rho_{n0}\right) \stackrel{d}{\longrightarrow} N(0, 1). 
\end{align*}
\end{Prop}

Proposition \ref{prop2} demonstrates that statistical inference for \(\hat\mu_n\), \(\hat\tau_n\), and \(\hat\rho_n\) can be conducted without knowledge of the sparsity level parameters. This allows for a unified approach to inference, even in cases where the limiting distribution of \(\hat{\rho}_n\) varies based on the relationship between \(a\) and \(b\) as seen in Corollary \ref{rho}.

\section{The $p_{1.5}$ Model}\label{Sec: Main results}
Having explored the statistical inference of the BR model in equation \eqref{eq:BR}, we now turn our attention to the \(p_{1.5}\) model presented in equation \eqref{eq:p1.5}. To gain deeper insights into this model, we focus on a fixed-dimensional covariate scenario, assuming that the dimensions of \(X_i\), \(Y_j\), and \(V_{ij}\) are all fixed. In this context, we consider their corresponding parameters as fixed quantities while allowing \(\mu_n\) and \(\rho_n\) to vary with \(n\), resulting in sparse networks. To address this sparsity, we define \(\mu_n = -a \log n + \mu\) and \(\tau_n := 2\mu_n + \rho_n = -b \log n + \tau\).

We assume that the dyads \(\{(A_{ij}, A_{ji})\}_{1 \leq i < j \leq n}\) are mutually independent, conditional on the covariates \(\{X_i\}_{i \geq 1}\), \(\{Y_j\}_{j \geq 1}\), and \(\{V_{ij}\}_{i \neq j}\). It is important to note that our model encompasses two sources of dependence. The first source arises from the covariates: two edges are dependent if they share at least one node (e.g., node \(k\)) due to the influence of \(X_k\) and \(Y_k\). The second source of dependence is the dyadic dependence, where \(A_{ij}\) and \(A_{ji}\) are dependent even when conditioned on \(\{X_i\}_{i \geq 1}\), \(\{Y_j\}_{j \geq 1}\), and \(\{V_{ij}\}_{i \neq j}\). We make the following assumptions:
\begin{Assum}{\label{ass2}} 
Assume that the true value \((\gamma_{10}^{T}, \gamma_{20}^{T}, \delta_0)^{T}\) of \((\gamma_{1}^{T}, \gamma_{2}^{T}, \delta)^{T}\) lies in the interior of \(\Omega_2\), where \(\Omega_2\) is a compact subset of \(\mathbb{R}^{d_1 + d_2 + d_3}\).
\end{Assum}

{\begin{Assum}{\label{ass3}} 
The covariates \(X_{i}\), \(Y_{j}\), and \(V_{ij}\) are centered, identically distributed, and uniformly bounded. For distinct pairs $\{i,j\}$ and $\{k,l\}$ where $\{i,j\} \cap \{k,l\} = \emptyset$, the covariates $\{X_{i}, Y_{i}, V_{ij}\}$ is independent of $\{X_{k}, Y_{k}, V_{kl}\}$. Moreover, the covariance matrix of the vector \((Z^{T}_{ij}+Z^{T}_{ji}, V^{T}_{ij})\), where $Z^{T}_{ij} = (X^{T}_{i}, Y^{T}_{j})$, is strictly positive definite.
\end{Assum}}

{Assumption \ref{ass2} is standard in the literature, whereas Assumption \ref{ass3} permits dependence among $X_i$, $Y_i$, and $V_{ij}$ while requiring independence across dyads}.  Let \(\theta = \left(\mu, \tau, \gamma_1^{T}, \gamma_2^{T}, \delta^{T}\right)^{T}\) represent the parameter vector, which does not depend on \(n\). The corresponding negative log-likelihood is defined as follows:
\[
\ell_{n}(\theta) = \sum_{i<j} \log(k_{i, j}(\theta)) - \sum_{i < j} \left(A_{ij} f^{(1)}_{ij}(\theta) + A_{ji} f^{(2)}_{ij}(\theta) + A_{ij} A_{ji} \left(f^{(3)}_{ij}(\theta) - f^{(1)}_{ij}(\theta) - f^{(2)}_{ij}(\theta)\right)\right),
\]
where 
$
f^{(1)}_{ij}(\theta) = \mu + X_{i}^{T} \gamma_1 + Y_{j}^{T} \gamma_2,  
f^{(2)}_{ij}(\theta) = \mu + X_{j}^{T} \gamma_1 + Y_{i}^{T} \gamma_2, 
f^{(3)}_{ij}(\theta) = \tau + \left(X_{i}^{T} + X_{j}^{T}\right) \gamma_1 + \left(Y_{i}^{T} + Y_{j}^{T}\right) \gamma_2 + V_{ij}^{T} \delta,
$
and 
$
k_{ij}(\theta) = 1 + n^{-a} \exp(f^{(1)}_{ij}(\theta)) + n^{-a} \exp(f^{(2)}_{ij}(\theta)) + n^{-b} \exp(f^{(3)}_{ij}(\theta)).
$ The maximum likelihood estimator \(\hat{\theta}\) is defined as:
\[
\hat{\theta} = \operatorname{argmin}_{\theta \in \Omega_1 \times \Omega_2} {\binom{n}{2}}^{-1} \ell_{n}(\theta).
\]

Let \(\eta = (\gamma^{T}_1, \gamma^{T}_2)^{T}, \phi = (\tau, \delta^{T})^{T}\). For the asymptotic properties of \(\hat{\theta}\), we have the following theorem:

\begin{theorem}\label{theorem1}
Under Assumptions \ref{ass1}, \ref{ass2}, and \ref{ass3}, as \(n\) approaches infinity, \(\hat{\theta}\) is consistent and asymptotically normal. Specifically:
\[
\left(\sqrt{n^{2-a}}(\hat{\mu}-\mu_0), \sqrt{n^{2-\min\{a,b\}}}(\hat{\eta}-\eta_0)^{T}, \sqrt{n^{2-b}}(\hat{\phi}-\phi_0)^{T}\right)^T \xrightarrow{d} N\left(0, 2H^{-1}(\theta)\right)
\]
where $H(\theta)$ is independent of \(n\) but varies across different scenarios of $a$ and $b$. The expression of $H(\theta)$ for each scenario is provided in the Supplementary Materials. 
\end{theorem}

Theorem \ref{theorem1} illustrates how the sparsity level of the network influences the rate of convergence for various parameters. Notably, the effective sample size for inferring \(\delta\) is given by \(n^{2-b}\), which depends on the configuration \((1,1)\) or the number of reciprocal links. Additionally, the convergence rate of \(\hat{\eta}\) is affected by the relative values of \(a\) and \(b\). This variation can be attributed to the presence of the nodewise covariates \(X_i\) and \(Y_j\) in both configurations \((1,0)\) and \((1,1)\). The expected total number of these configurations is on the order of \(O(n^{2-\min\{a,b\}})\), thus highlighting the dependence on the relative values of \(a\) and \(b\). {Further examination of the structure of $H(\theta)$ in the Supplementary Materials reveals the following insight.  Notably, when $a=b$, $\sqrt{n^{2-\min\{a,b\}}}\hat{\eta}$ is correlated with both $\sqrt{n^{2-a}}\hat{\mu}$ and $\sqrt{n^{2-b}}\hat{\phi}$. However, it is correlated only with $\sqrt{n^{2-a}}\hat{\mu}$ when $a<b$ and only with $\sqrt{n^{2-b}}\hat{\phi}$ when $a>b$. Remarkably these rates are \textit{explicit} for arbitrarily sparse networks as long as $a,b>0$. In contrast, for the directed network models considered in  \cite{yan2016asymptotics} and \cite{yan2019statistical}, which do not account for reciprocity, only \textit{implicit} rates of convergence have been established for relatively dense networks.}

As with the BR model, the sparsity indices $a$ and $b$ are typically unknown. Let \(\theta_n = \left(\mu_n, \gamma_1^{T}, \gamma_2^{T}, \tau_n, \delta^{T}\right)^{T}\) represent the parameter vector, and \(\hat{\theta}_n\) be the maximum likelihood estimator (MLE) of \(\ell_{n}(\theta_n)\), the likelihood function defined with \(\theta_n\) as its parameters. The following proposition presents a practical approach for conducting statistical inference without requiring knowledge of the network's sparsity levels.

\begin{Prop}\label{prop3}
Under Assumptions \ref{ass1}, \ref{ass2}, and \ref{ass3}, as \(n \to \infty\), with probability approaching one, \(\hat{\theta}_n\) exists. For \(k = 1, \ldots, d_1 + d_2\), we have the following result:
\[
n \cdot \frac{\hat{\eta}_k - \eta_{0k}}{\sqrt{2\left(H_n^{-1}(\hat{\theta}_n)\right)_{1+k, 1+k}}} \stackrel{d}{\longrightarrow} N(0, 1),
\]
where \(H_n(\hat{\theta}_n)\) denotes the Hessian matrix of \(\frac{1}{\binom{n}{2}}\ell_n(\hat{\theta}_n)\). Similar results can be derived for the other components of \(\hat{\theta}_n\), though we omit these details for brevity.
\end{Prop}

Proposition \ref{prop3} shows that statistical inference for \( \theta_n \) can be conducted within a unified framework, regardless of the network's sparsity levels or the specific forms of \( H(\theta) \) across different cases. It also provides implicit convergence rates for each parameter estimator. For instance, the rate for \(\hat{\eta}_k\) is expressed as 
\[
\frac{n}{\sqrt{2\left(H_n^{-1}(\hat{\theta}_n)\right)_{1+k, 1+k}}}.
\]
However, its precise dependence on $n$ remains unclear, as $H_n(\hat\theta_n)$ depends on $n$ and does not converge to a deterministic matrix.

\section{Simulation}\label{Sec: Simulation}

\begin{figure}[h!]
    \centering
\includegraphics[width=.90\textwidth]{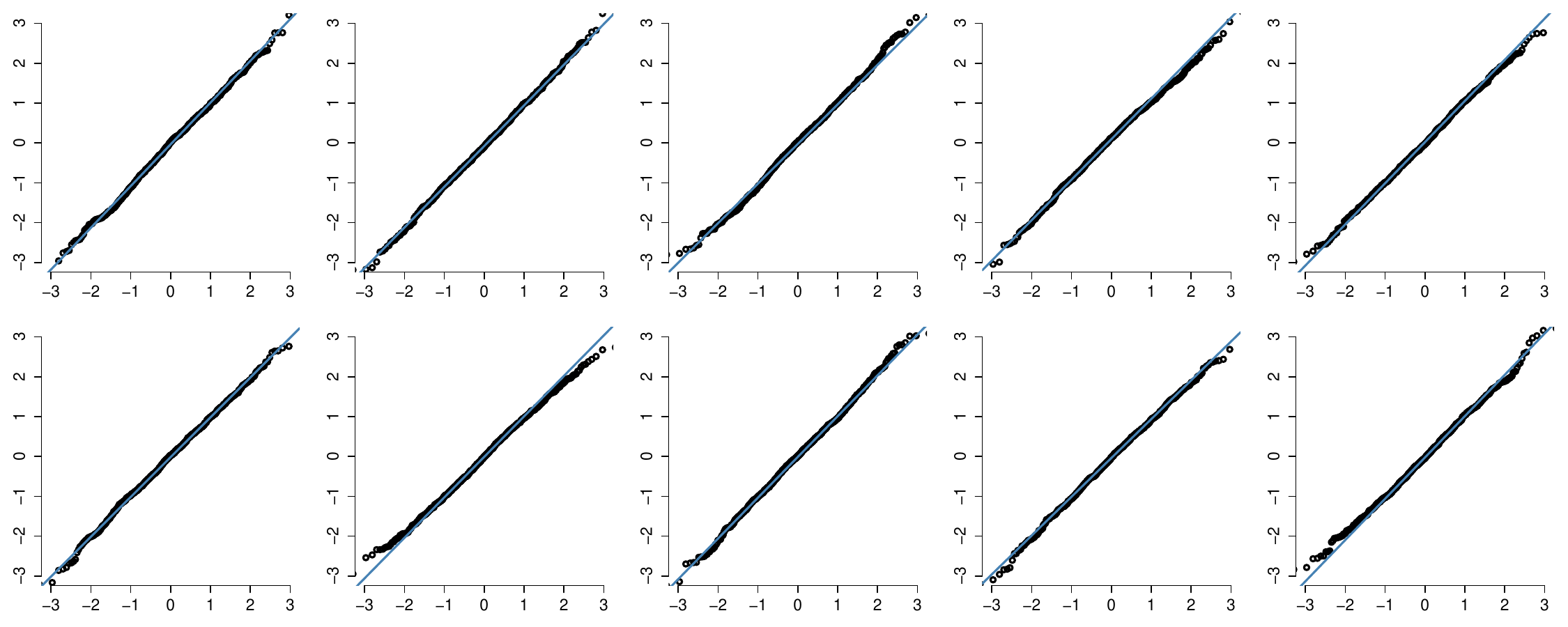}    
    \begin{picture}(0, 0)
        \put(-40, 175){\makebox(0, 0)[b]{\small $\hat{\delta}$}}
        \put(-130, 175){\makebox(0, 0)[b]{\small $\hat{\gamma}_2$}}
         \put(-220, 175){\makebox(0, 0)[b]{\small $\hat{\gamma}_1$}}
        \put(-310, 175){\makebox(0, 0)[b]{\small $\hat{\tau}$}}
         \put(-400, 175){\makebox(0, 0)[b]{\small $\hat{\mu}$}}
    \end{picture}
    \caption{The QQ plots of the standardized estimators for $a = b = 0.5$ are shown for $n=200$ (first row) and $n=1000$ (second row). The $x$-axis represents the theoretical quantiles, while the $y$-axis displays the sample quantiles.}
\label{figure2}
\end{figure}
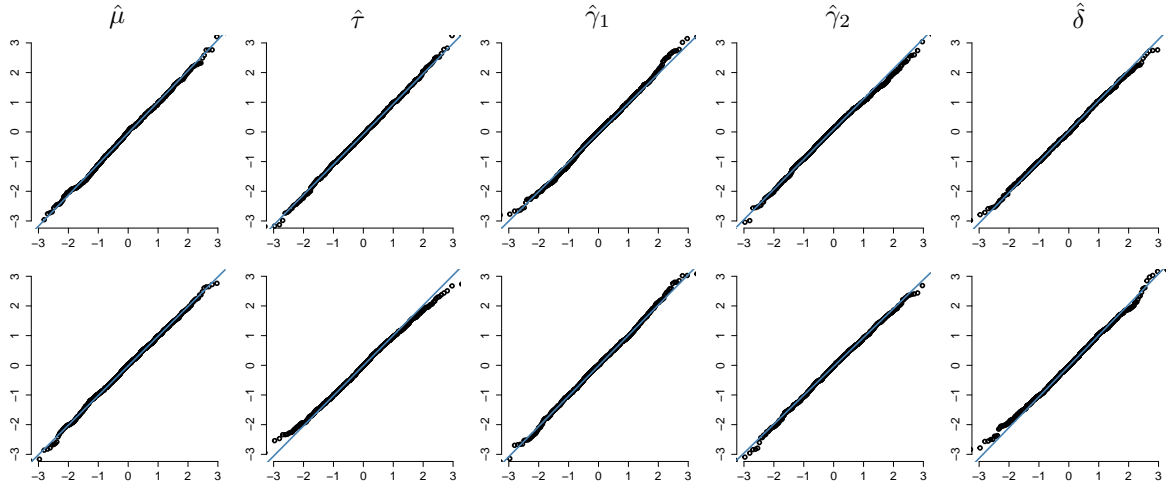

We evaluate the finite sample performance of our maximum likelihood estimator (MLE) for the \(p_{1.5}\) model through simulations. In our experiments, we set \(n = 200 \) or \(1000\), \(a = b = 0.5\), and the true parameter vector \(\theta_0 = (\mu_{0}, \tau_{0}, \gamma_{10}, \gamma_{20}, \delta_{0})^T = (0.2, 0.5, 0.2, 0.4, 0.3)^T\). We generate the covariates \(\{X_{i}\}_{i \geq 1}\), \(\{Y_{j}\}_{j \geq 1}\), and \(\{V_{ij}\}_{i \neq j}\) from the standard uniform distribution. 
We generate a total of \(1000\) networks from this model and evaluate the asymptotic normality of the MLEs using quantile-quantile (QQ) plots. The results, illustrated in Figure \ref{figure2}, clearly indicate that these MLEs conform to normal distributions. 

Next, we examine the normal approximation for our estimator using Proposition \ref{prop3}. In the following table, we keep \(\theta_0\) and the generating scheme for \(\{X_{i}\}_{i \geq 1}\), \(\{Y_{j}\}_{j \geq 1}\), and \(\{V_{ij}\}_{i \neq j}\) fixed while varying the sparsity parameters \(a\) and \(b\). 
Table \ref{Table: inference gamma} presents the results for \(\gamma_1\), as the findings for the other parameters are comparable. The coverage of the confidence intervals is consistently close to the \(95\%\) level across all network sizes and sparsity levels. This empirically supports the validity of the asymptotic results derived in Theorem \ref{theorem1}. Additionally, we observe that the median length of the confidence interval decreases with increasing network size, which is expected; as the network becomes sparser, the effective sample size decreases.

\begin{table}[!htbp]
    \centering
    \begin{tabular}[t]{rrrrrrrrrrrrr}
        \toprule
            $n$ & \multicolumn{2}{c}{{$a=0.5$, $b=0.5$}} && \multicolumn{2}{c}{{$a=1$, $b=0.5$}} && \multicolumn{2}{c}{{$a=1$, $b=1$}} && \multicolumn{2}{c}{{$a=1$, $b=1.5$}}\\
        \midrule
        & Coverage & Width && Coverage & Width && Coverage & Width && Coverage & Width \\
        \midrule
        \addlinespace[0.3em]
        $200$  & $94.2\%$ & $0.094$ && $94.3\%$ & $0.109$ && $95.6\%$ & $0.284$ && $94.6\%$ & $0.417$ \\
        $500$  & $94.9\%$ & $0.043$ && $95.8\%$ & $0.052$ && $93.9\%$ & $0.177$ && $94.7\%$ & $0.265$ \\
        $800$  & $94.7\%$ & $0.030$ && $95.3\%$ & $0.036$ && $95.1\%$ & $0.139$ && $94.8\%$ & $0.210$ \\
        $1000$ & $93.8\%$ & $0.025$ && $95.5\%$ & $0.030$ && $96.7\%$ & $0.124$ && $95.8\%$ & $0.189$ \\
        \bottomrule
    \end{tabular}
    \caption{Empirical coverage under nominal 95\% coverage and median lengths of confidence intervals for $\gamma_1$. The results are similar for the other components of $\theta_n$.}
    \label{Table: inference gamma}
\end{table}

\section{Data Analysis}\label{Sec: Data Analysis}
We further illustrate our results by applying our method to two real-world datasets.

\noindent
\textbf{Analysis of Lazega's Dataset.} We begin with an analysis of Lazega's dataset of lawyers \citep{lawyernetwork}, which has also been examined in studies such as \cite{yan2019statistical}. This dataset captures the interactions among 71 lawyers (36 partners and 35 associates) at a New England law firm. Our focus is on the basic advice network, where a directed edge from lawyer \(i\) to lawyer \(j\) indicates that lawyer \(i\) has sought basic professional advice from lawyer \(j\). This directed network has a density of 0.18, with in-degrees ranging from 0 to 30 and out-degrees from 0 to 37. In addition to the network structure, several covariates for each lawyer were collected, including their status (partner or associate), gender (male or female), office location (Boston, Hartford, or Providence), years of tenure with the firm, age, practice area (litigation or corporate), and the law school attended (Harvard, Yale, UConn, or others).

\begin{table}[!htp]
	\centering
	\begin{tabular}{clrr}
		\toprule
		Type & Covariate & Estimate & Confidence Interval\\
                 \midrule
		$X$ & Age & $-0.03$ & $(-0.04, -0.02)$ \\
		$Y$&Years with firm & $0.05$ & $(0.04, 0.06)$\\
		$V$&Same status & $1.64$ & $(1.23, 2.04)$ \\
		&Same office & $1.67$ & $(1.25, 2.08)$\\
		&Same practice & $1.32$ & $(0.95, 1.69)$\\
		&Same gender & $0.31$ & $(-0.07, 0.68)$ \\
		&Same law school & $0.11$ & $(-0.23, 0.46)$\\
		\bottomrule
	\end{tabular}
	\caption{\label{tab:lawyer results} Estimation for Lazega's Lawyer friendship network and $95\%$ confidence intervals.}
\end{table}
For the covariates \(X_i\) and \(Y_j\), we utilize the nodewise variables of age and years with the firm, respectively. For the dyad covariate \(V_{ij}\), we employ an indicator that denotes whether the lawyers share the same status, office location, practice area, gender, or law school attended. We fitted our model, and the resulting estimates along with their \(95\%\) confidence intervals are presented in Table \ref{tab:lawyer results}. 
The findings in this table align well with our expectations regarding the signs of the estimates. Specifically, younger lawyers tend to seek advice more frequently, as indicated by the negative estimate for age, while those with more years at the firm are more likely to be consulted, reflected in the positive estimate for years with the firm. Additionally, lawyers are inclined to seek professional advice from colleagues who share the same office, status, or practice area. 
However, while our point estimates for sharing the same gender and attending the same law school are positive, their confidence intervals extend into the negative range. Consequently, we cannot draw definitive conclusions about the influence of these factors on network formation.

\noindent
\textbf{Analysis of the Trade Partnerships Network.} We now turn our attention to the trade partnerships network data collected by \cite{Silva:Tenreyro:2006} and analyzed by \cite{stein2024sparse}. This dataset encompasses a cross-section of $136$ countries and their bilateral export flows in $1990$. It also includes valuable attributes for each country, such as real GDP per capita, tariff rates, and whether a country is landlocked, along with various dyadic characteristics that measure the closeness between each pair of countries. These characteristics include factors like distance, the presence of a common language, and whether there is a free trade agreement between the countries. 
In our analysis, we establish a directed edge from country \(i\) to country \(j\) if the trading volume between them accounts for at least \(1\%\) of country \(i\)'s total trading volume. This indicates that country \(j\) is a significant trade partner for country \(i\). The resulting directed network comprises $136$ nodes and $2,141$ edges, resulting in an edge density of \(11.7\%\). Additionally, there are $260$ mutual edges in this network, indicating that $260$ pairs of countries have important trade partnerships with each other.

\begin{table}[!htbp]
	\centering
	\begin{tabular}{clrr}
		\toprule
		Type&Covariate & Estimate & Confidence Interval \\
                 		\midrule
		$X$&Log GDP & $1.13$ & $(1.08, 1.17)$\\
		$Y$&Landlocked & $-0.12$ & $(-0.29, 0.05)$\\
		$V$&Log distance & $-1.62$ & $(-1.84, -1.41)$\\
		&Common border & $1.70$ & $(1.06, 2.35)$ \\
		&Common language & $1.67$ & $(1.15, 2.19)$ \\
		&Colonial ties & $0.93$  & $(0.37, 1.49)$ \\
		&Preferential trade agreement & $0.84$  & $(0.17, 1.51)$\\
		\bottomrule
	\end{tabular}
	\caption{\label{tab:trade} Estimation for world trade data and $95\%$ confidence intervals.}
\end{table}

For the dyad covariates \(V_{ij}\), we adopt the same approach as \cite{jochmans}. Specifically, we include the logarithm of the geographical distance between the capitals of countries \(i\) and \(j\), along with several dummy variables that indicate the presence of specific relationships between the countries. These dummy variables include common border, common language, colonial ties, and preferential trade agreements. For the nodewise covariate \(X_i\), we utilize a dummy variable indicating whether a country is landlocked, as landlocked countries generally have less active international trade. For the covariate \(Y_j\), we select the log of GDP per capita, based on the belief that well-developed countries are more likely to attract important trade partnerships with others.

We fitted our model to the network using the covariates described above, and the results are summarized in Table \ref{tab:trade}. Our findings indicate that a higher GDP per capita significantly enhances a country's attractiveness to other nations. In contrast, the confidence interval for the landlocked coefficient includes zero, suggesting insufficient evidence to confirm a meaningful impact of landlockedness on international trade partnerships. Additionally, the results for the dyad covariates align with our expectations: having a preferential trade agreement, speaking the same language, sharing a border, or having colonial ties all positively influence trade partnerships between countries, while greater geographical distance exerts a strong negative effect.

\renewcommand{\baselinestretch}{1}
\normalsize
 \bibliographystyle{agsm}
\bibliography{reference}
 
\end{document}